\documentclass[epj]{svjour}
\usepackage{graphics}
\usepackage{epsfig}
\def\ra{\rightarrow}
\begin{document}
\title{ECFA-Summary}
\subtitle{Higgs, gamma-gamma and  e-gamma physics
}
\author{Maria Krawczyk\inst{1} 
}                     
%
%
\institute{Institute of Theoretical Physics, Warsaw University, Ho\.za 69,
00-681 Warsaw, Poland  
}
\date{Received: date / Revised version: date}
\abstract{
Recent results obtained within ECFA/DESY and ECFA Study by the Higgs and 
$\gamma \gamma/e \gamma$ physics working groups are presented.
} 
\maketitle
\section{Introduction}
\label{intro}
The recents results obtained within ECFA/DESY and
ECFA Study for a Linear Collider (LC)
for Higgs search in $e^+e^-$ mode and in $\gamma \gamma/ 
e \gamma$ option  (Photon Linear Collider - PLC) are presented. 
For $e \gamma$ option results for testing anomalous 
gauge couplings are also shown.
 The extensive summary of the studies of Higgs physics
in $e^+e^-$ collisions  and on
physics at PLC
can be found in \cite{Desch:2003xq} and \cite{DeRoeck:2003gv}, respectively.
\section{Higgs studies for an \protect{$e^+e^-$} Linear Collider}
\label{sec:1}
\begin{figure*}
\centering 
\resizebox{0.47\textwidth}{!}{%
\includegraphics{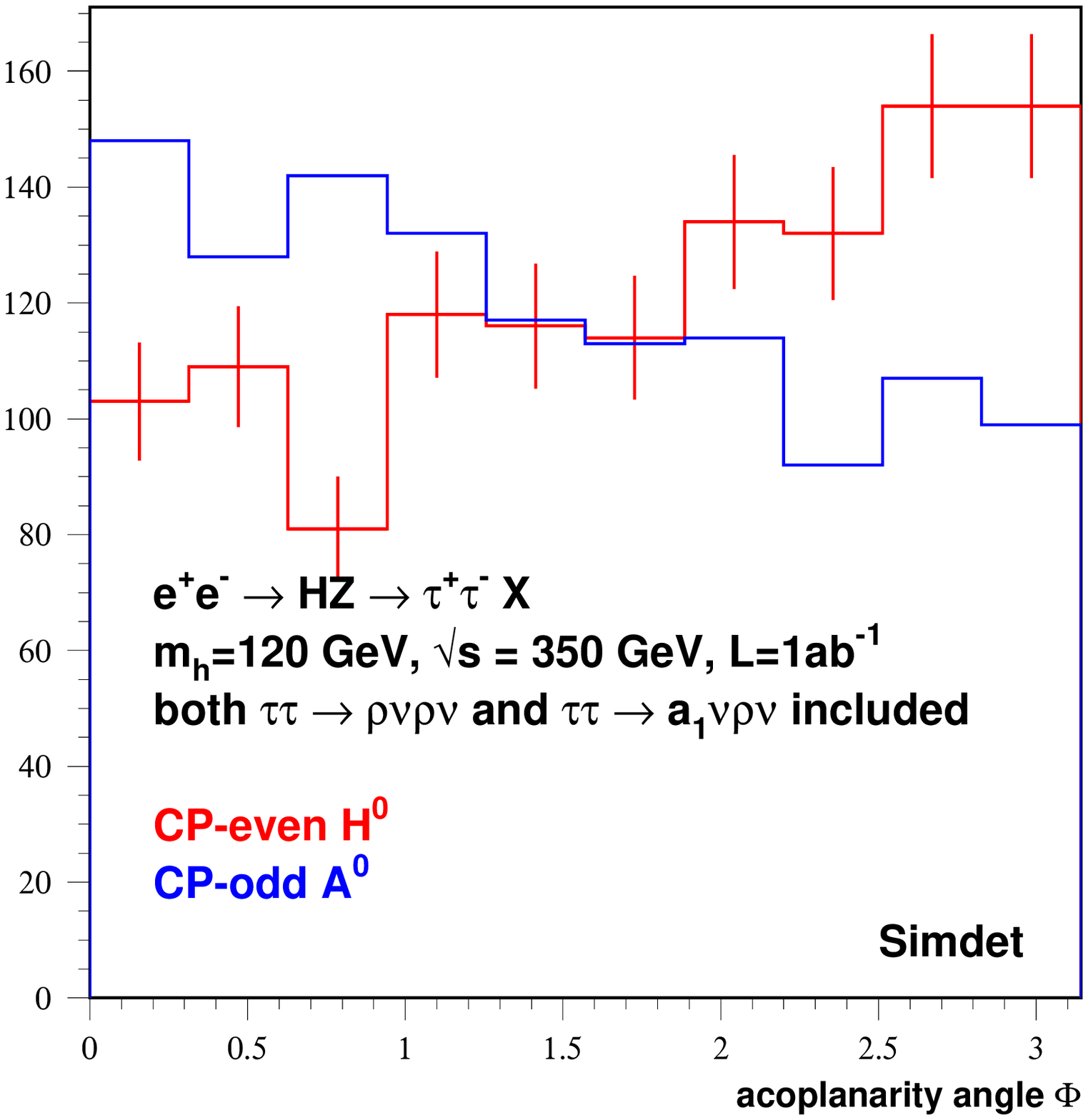}\includegraphics{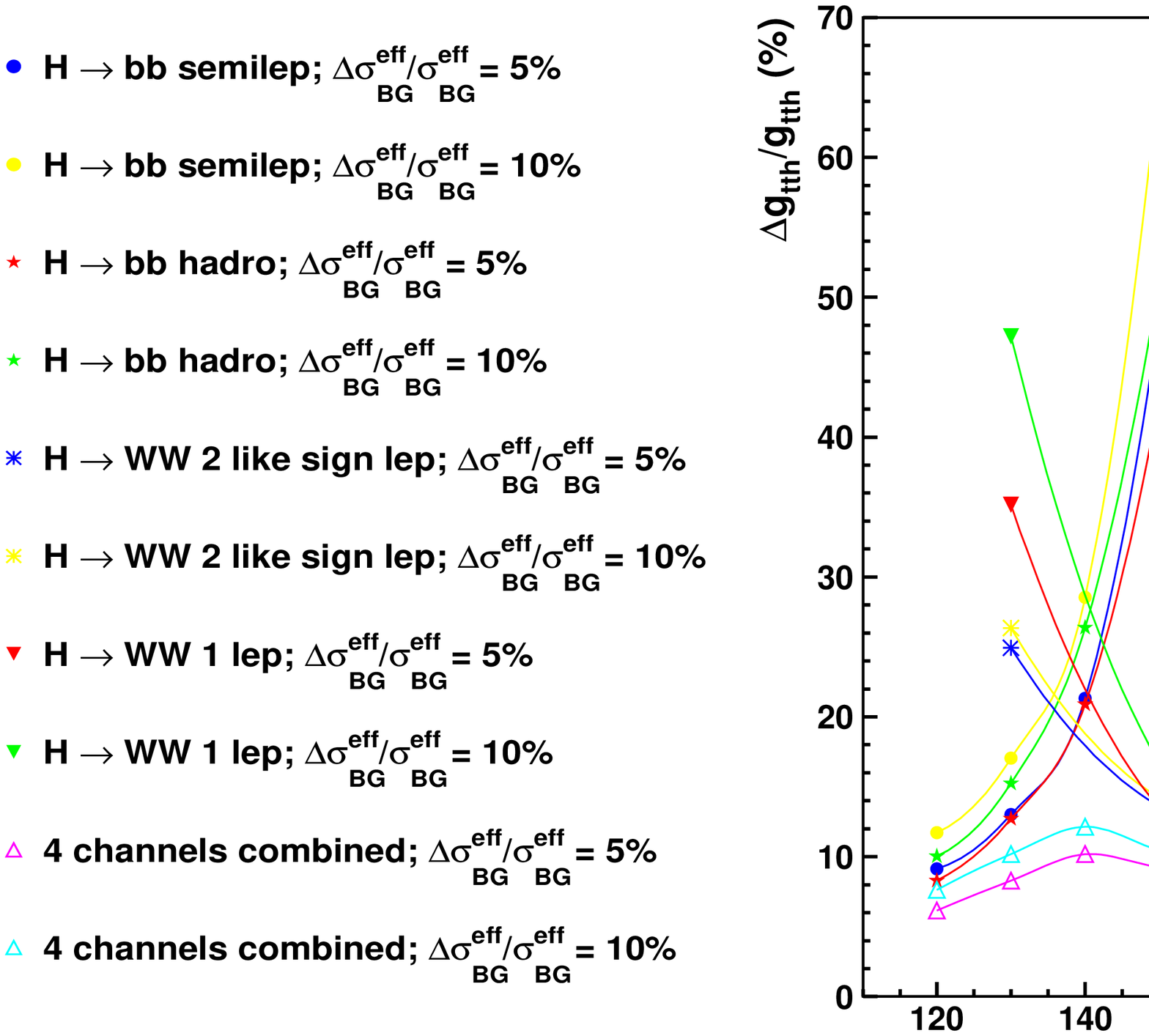}
}
\caption{Left: Distinguishing scalar from pseudoscalar using tau's; 
Right: Relative precision for $g_{tth}$ from various channels.}
\label{fig:1}       
\end{figure*}

\begin{figure*}
\centering
\resizebox{0.6\textwidth}{!}{%
\includegraphics{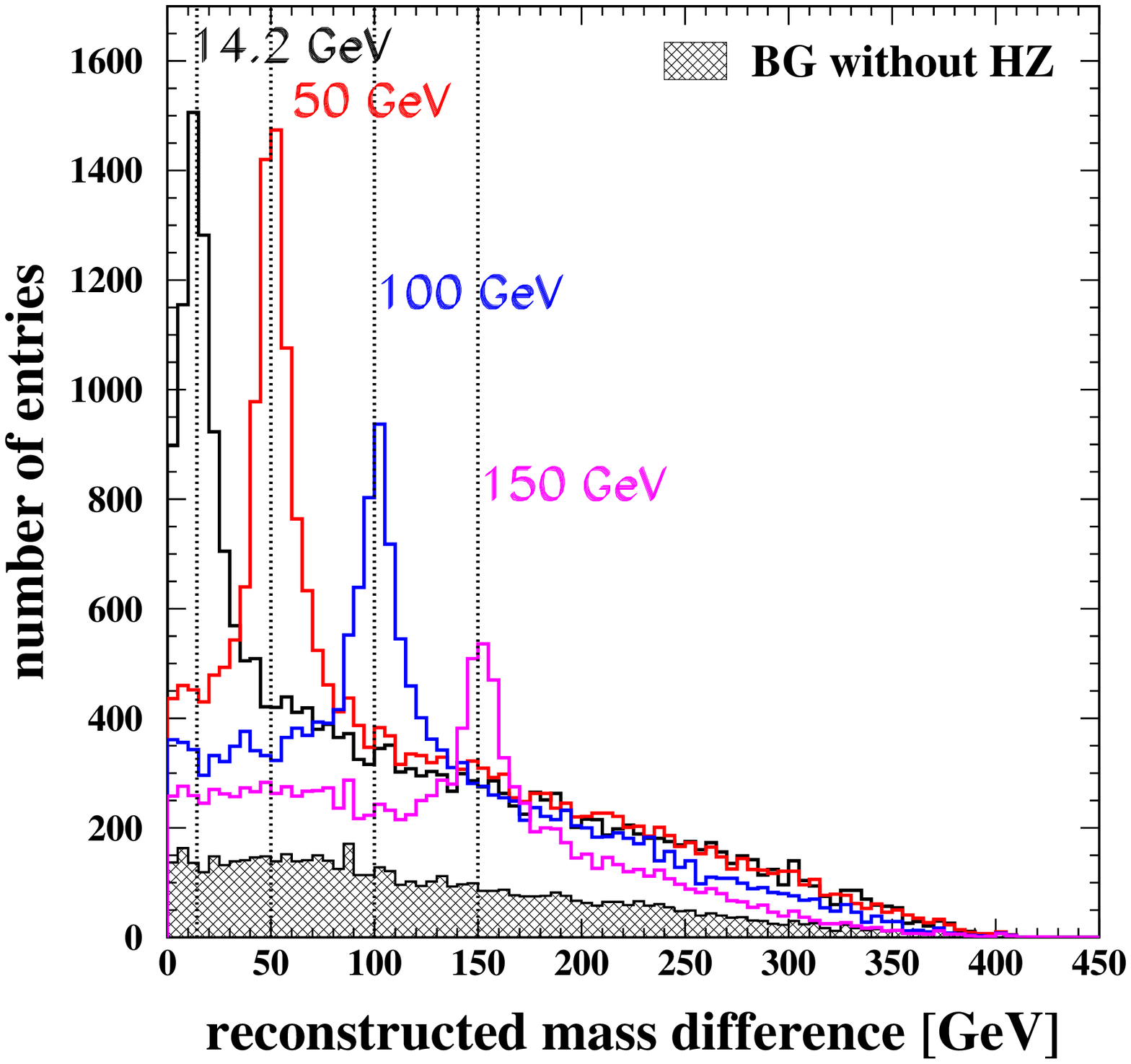}\hspace*{2cm}
\includegraphics{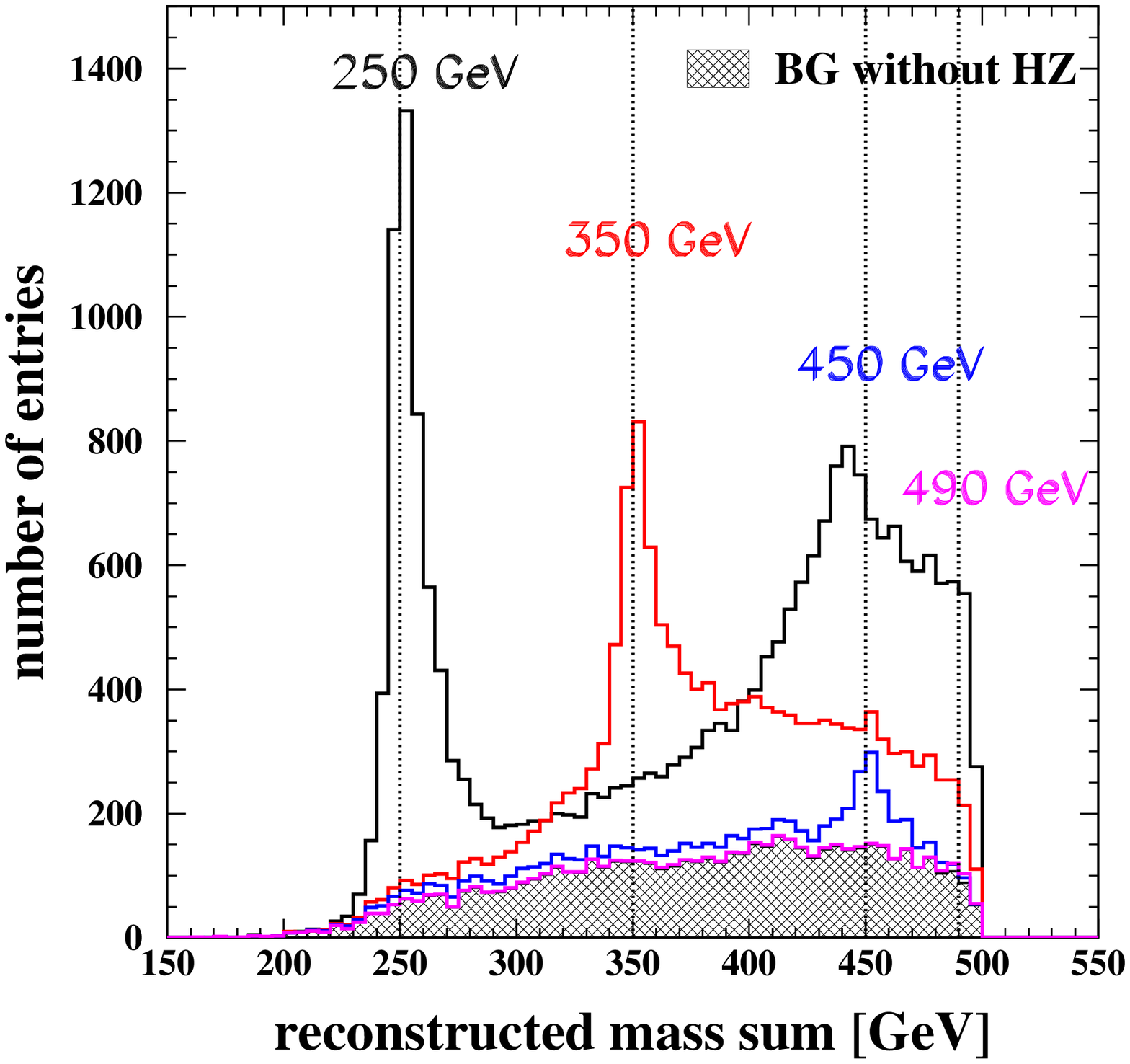}
}
\caption{ Results
for the reconstructed  difference and sum of mass of Higgs bosons in MSSM, 
for the $b\bar b$ final state.}
\label{fig:2}       
\end{figure*}

The Linear Collider is considered as a tool for precision Higgs measurements,
as it was shown in TESLA TDR \cite{tdr}. The further study was 
concentrated on  more realistic simulations of essential processes, 
and studying of new theoretical ideas and  
LHC-LC synergy.
 \paragraph{Higgs Quantum Numbers}
New ideas how to test  the spin and CP-parity of Higgs bosons were presented recently. 
One bases on   Higgs boson decay into $ ZZ$ \cite{mill} 
(results for PLC based on this idea are shown below). The other method uses  
the decay  $H \ra \tau \bar \tau $, with further decay of  tau's  into $\rho$, 
where the correlation of the decay products of $\tau$'s 
allows to establish the CP-parity of a Higgs boson. The study of 
the  process $e^+e^-\ra HZ \ra 
\tau \bar \tau X$ for CM energy equal to 350 GeV and luminosity 1 
$ab^{-1}$ \cite{worek}
shows that one can discriminate the scalar SM-Higgs with mass 120 GeV 
from the  pseudoscalar one (with the same production rate as for $H$)
at the 8 $\sigma$ level, see Fig.~\ref{fig:1} (Left).
\paragraph{Top Yukawa coupling}
New analysis \cite{top} 
of the measurement of the Yukawa coupling of the SM  Higgs
particle $h$ to 
top quarks is extended to higher masses, up to 200 GeV,
 with inclusion of the $h \ra WW$, and 
with full 6-fermion background (BG). 
The results for expected relative precision for $g_{tth}$ are presented in 
Fig.~\ref{fig:1} (Right), for the energy of 
collision of 800 GeV, luminosity of 1 $ ab^{-1}$ and
various final states 
(for two different background normalizations).
Combining  channels the precision  
can reach 6 to 14 \%.    
\paragraph{Supersymmetric Higgs Bosons}
The study of  heavy Higgs bosons  $H$ and $A$ 
has been performed for a particular  MSSM scenario \cite{susy}, in 
which the lighest Higgs boson $h$ couples to gauge bosons with a 
full strength ($\sin(\beta-\alpha)=1$). Then  $H$, with couplings 
to gauge bosons proportional to $\cos (\beta -\alpha)$, is produced
in $e^+e^-$ collision predominanty  in pair with $A$, 
with cross section $\propto \sin ^2(\beta -\alpha)$.
The decays of $H$ and $A$ are mainly to fermions $b$ and $\tau$'s,
and both $H$ and $A$ are nearly degenerate in masses.
The reconstructed  difference and sum of masses, 
for the $b\bar b$ final state, with $Br(H,A\ra b \bar b)=$0.9, 
presented in Fig.~\ref{fig:2} for energy of $e^+e^-$ collider
of 500 GeV with luminosity of 500 $fb^{-1}$ correspond to a precision 0.2 
to 2.8 GeV.
\section{Higgs resonance at  Photon Linear Collider}
\label{sec:2}
\begin{figure*}
\centering
\resizebox{0.95\textwidth}{!}{%
 \includegraphics{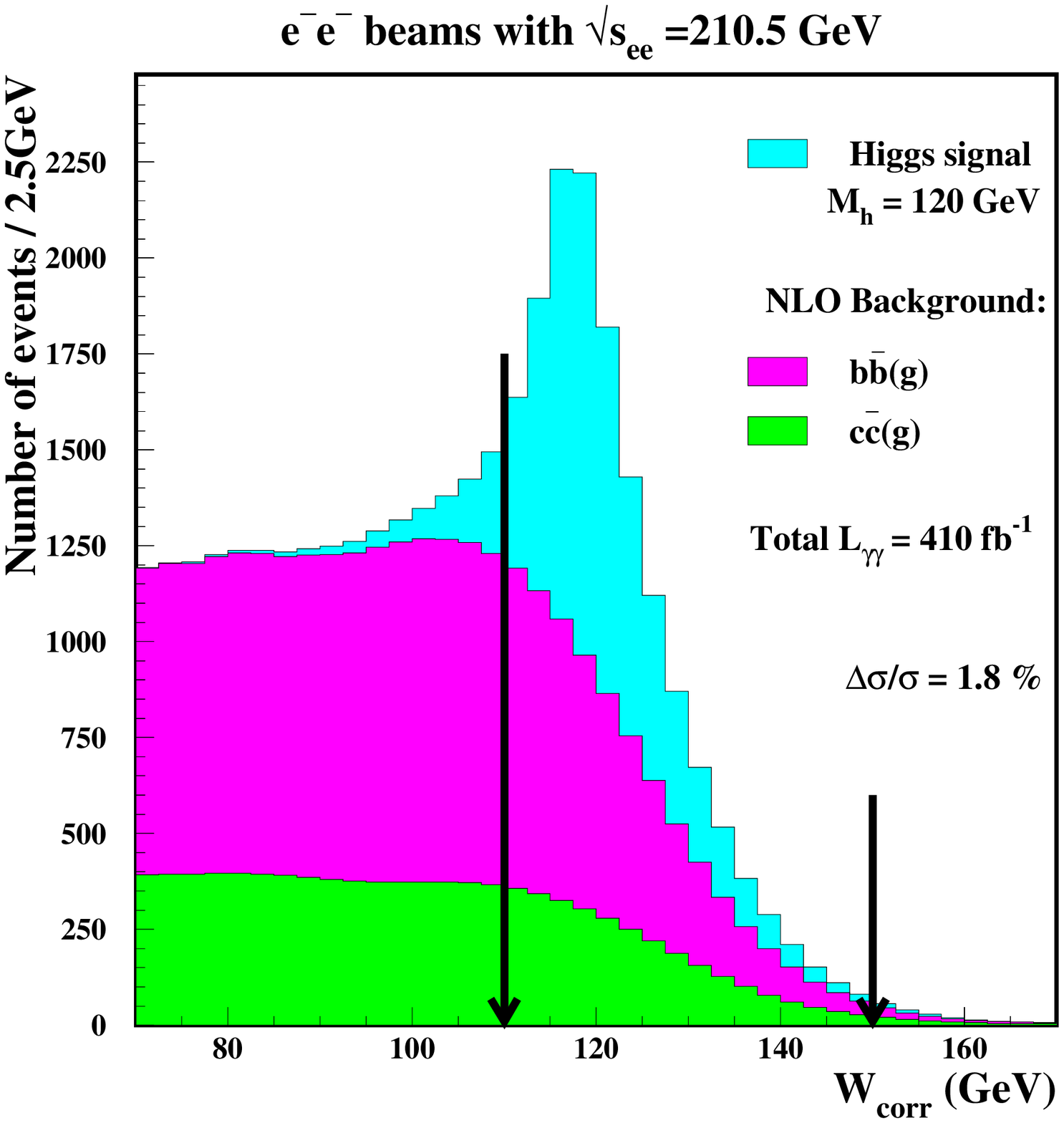}
\includegraphics{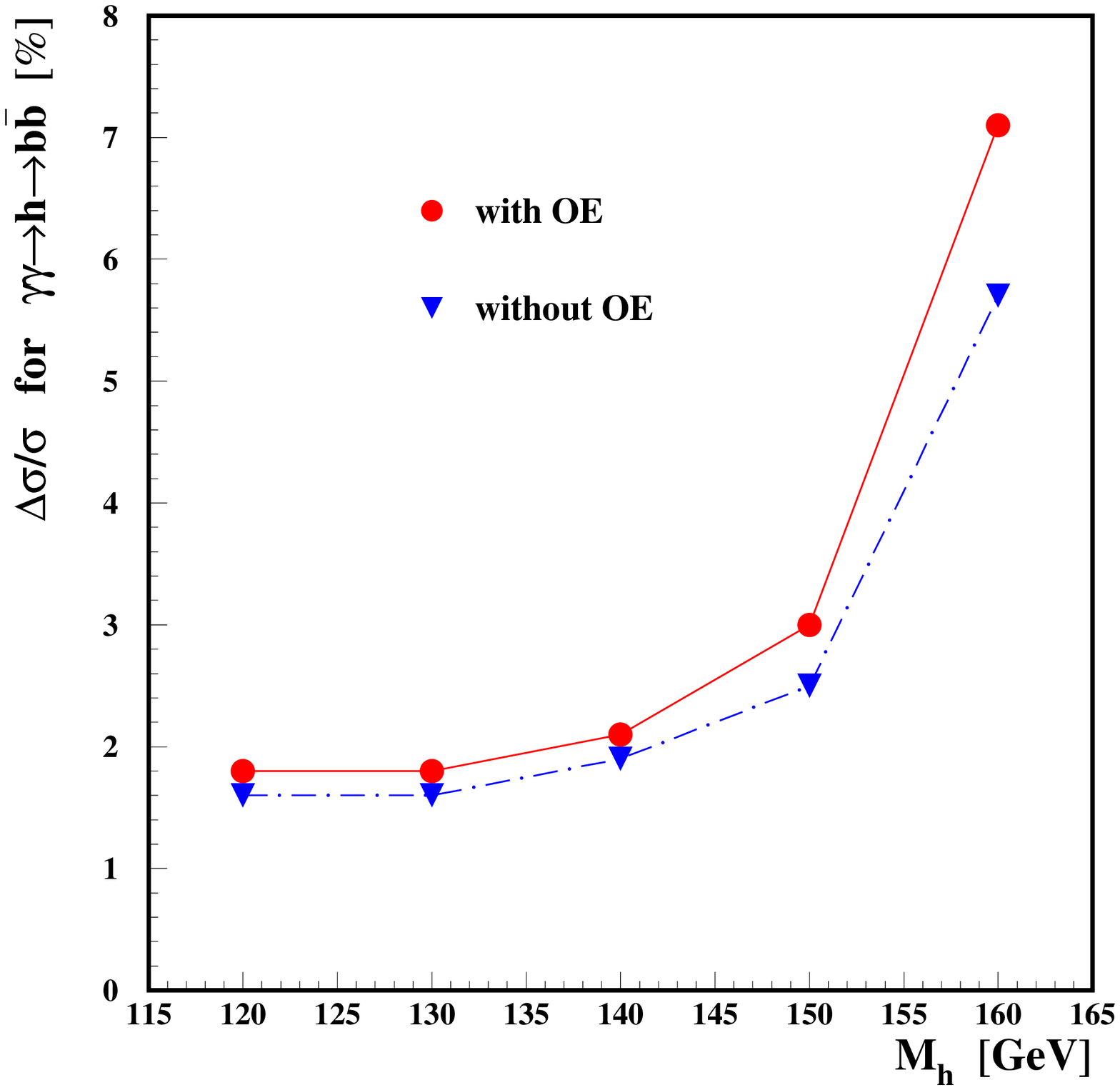}
\includegraphics{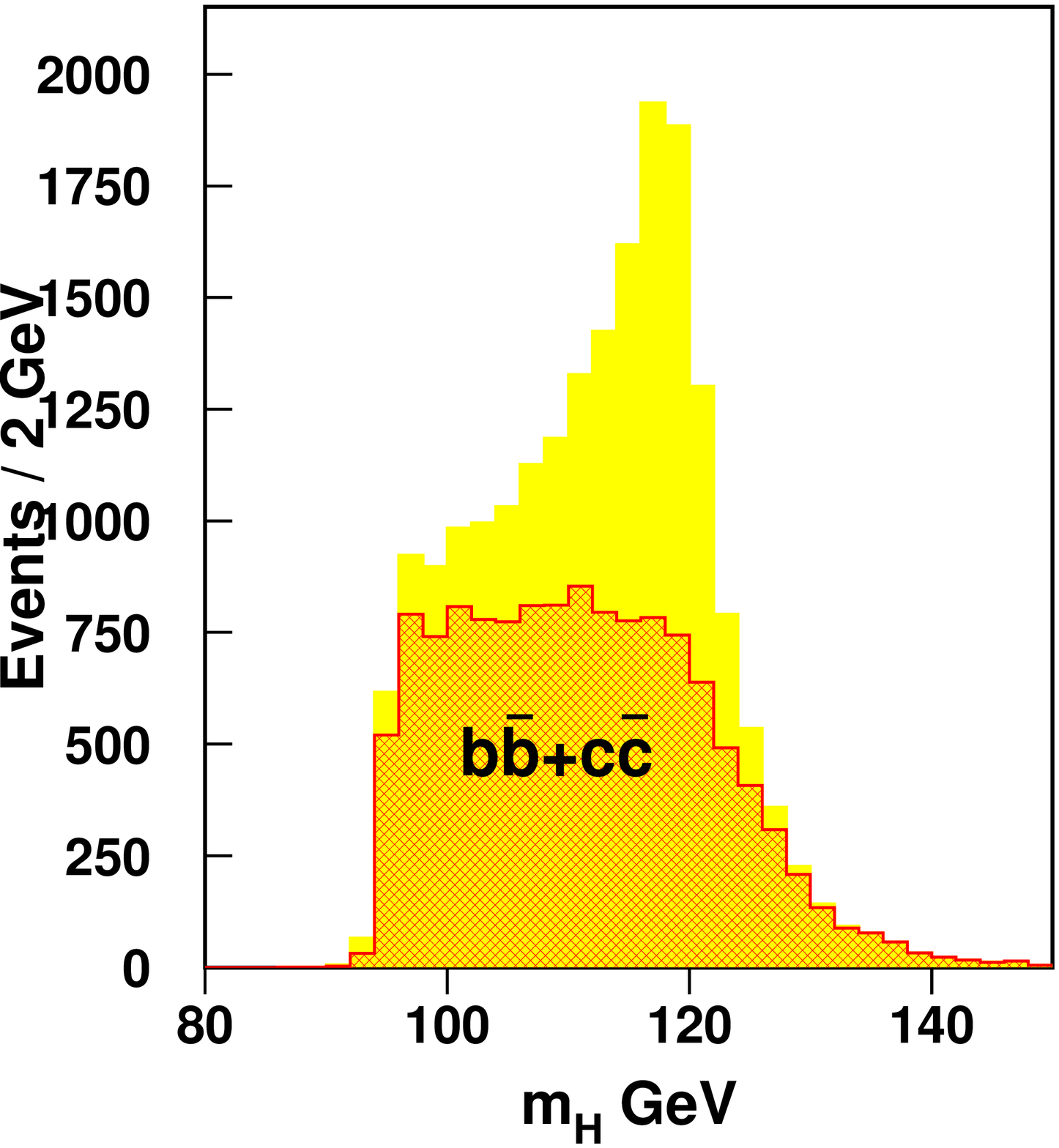}
}
\caption{Results for production of the SM Higgs with mass 120 GeV in 
$\gamma \gamma \ra h \ra b \bar b$ (Left: from \cite{Niezurawski:2003iu}, 
Right: from \cite{Rosca:2003wa}); Middle: A precision of measurement of the cross section
as a function of mass with and without  OE included in analysis \cite{Niezurawski:2003iu}.}
\label{fig:3}    
\end{figure*}

\begin{figure*}
\centering
\resizebox{0.95\textwidth}{!}{%
  \includegraphics{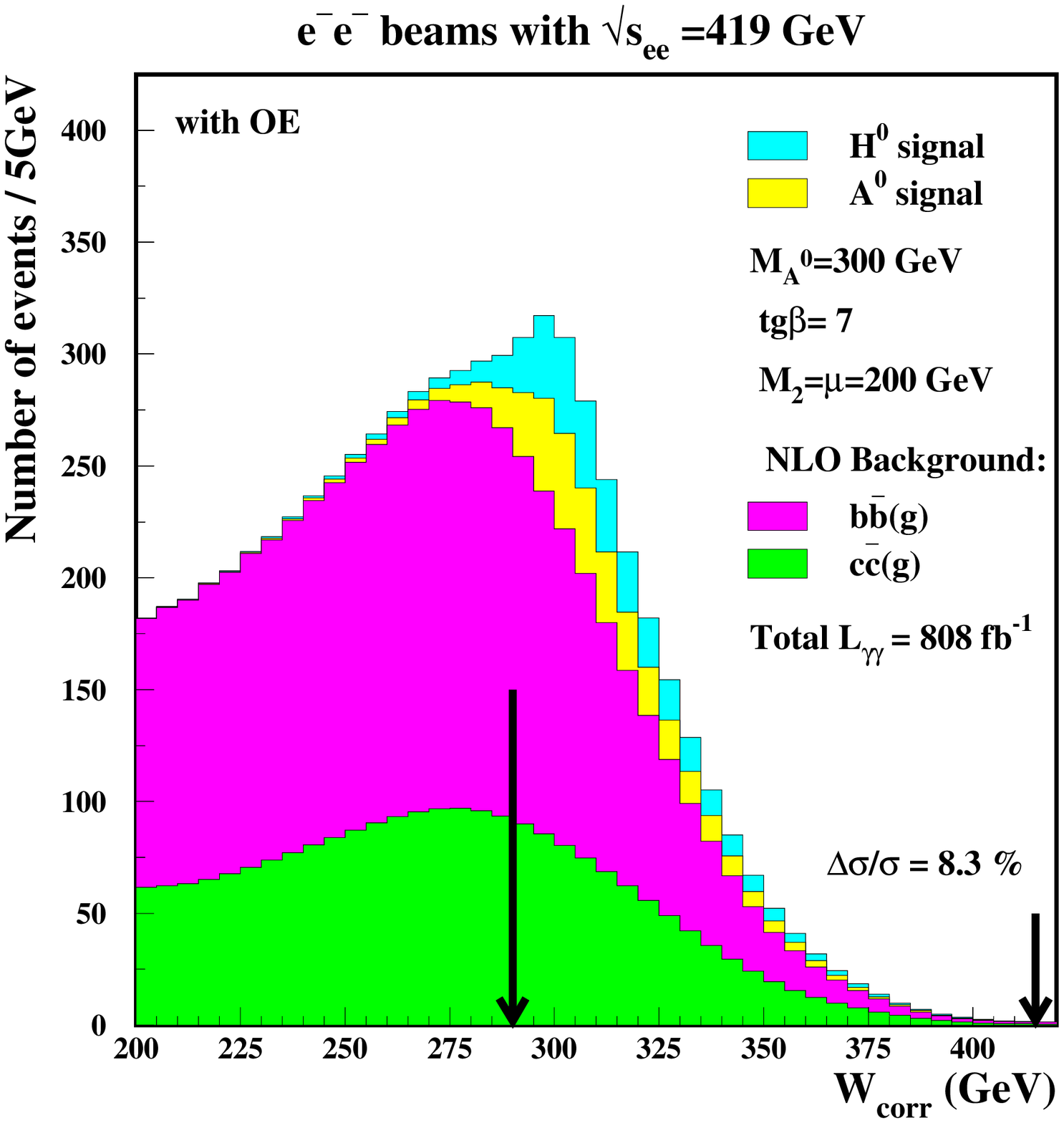} \includegraphics{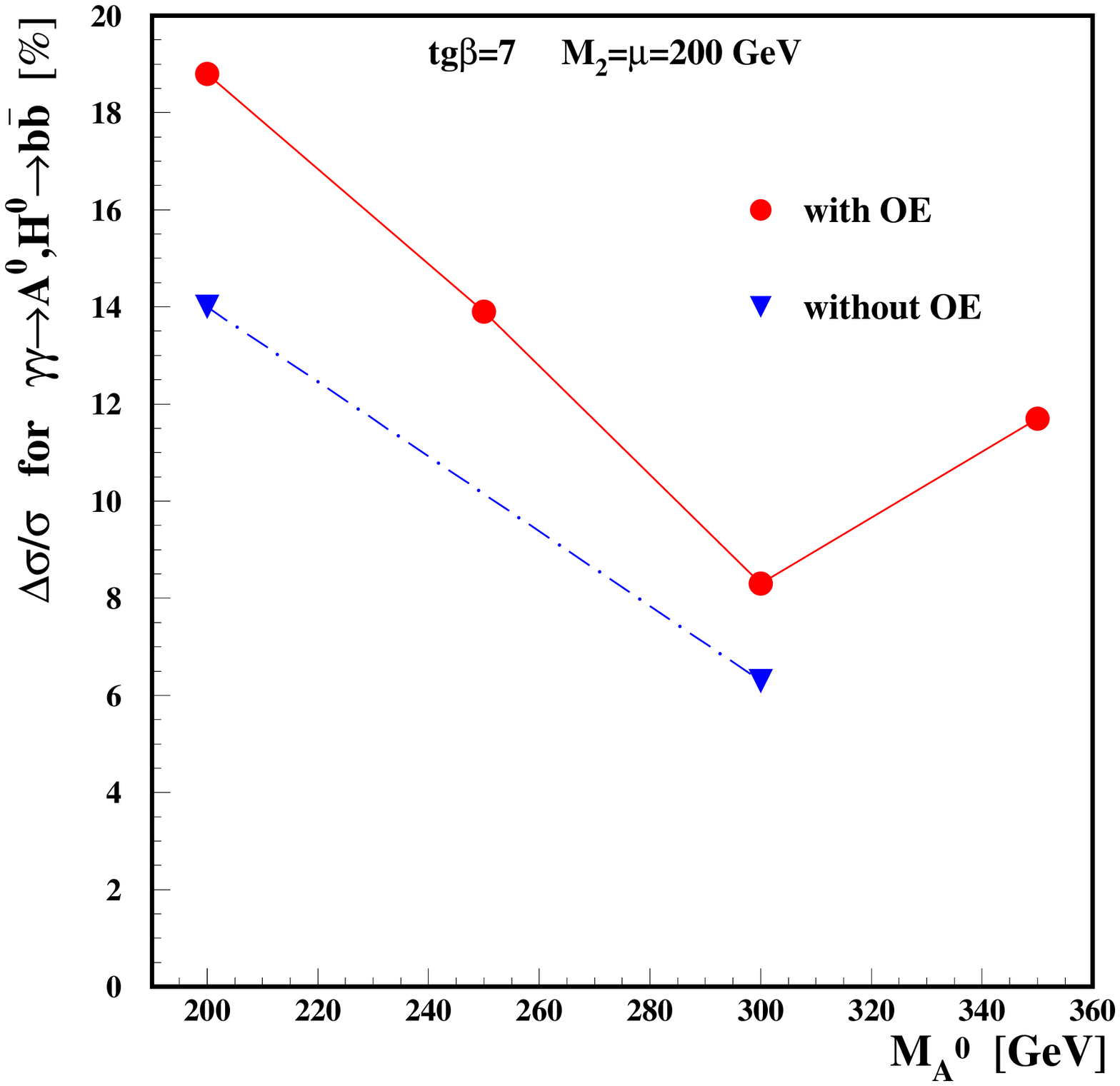}\includegraphics{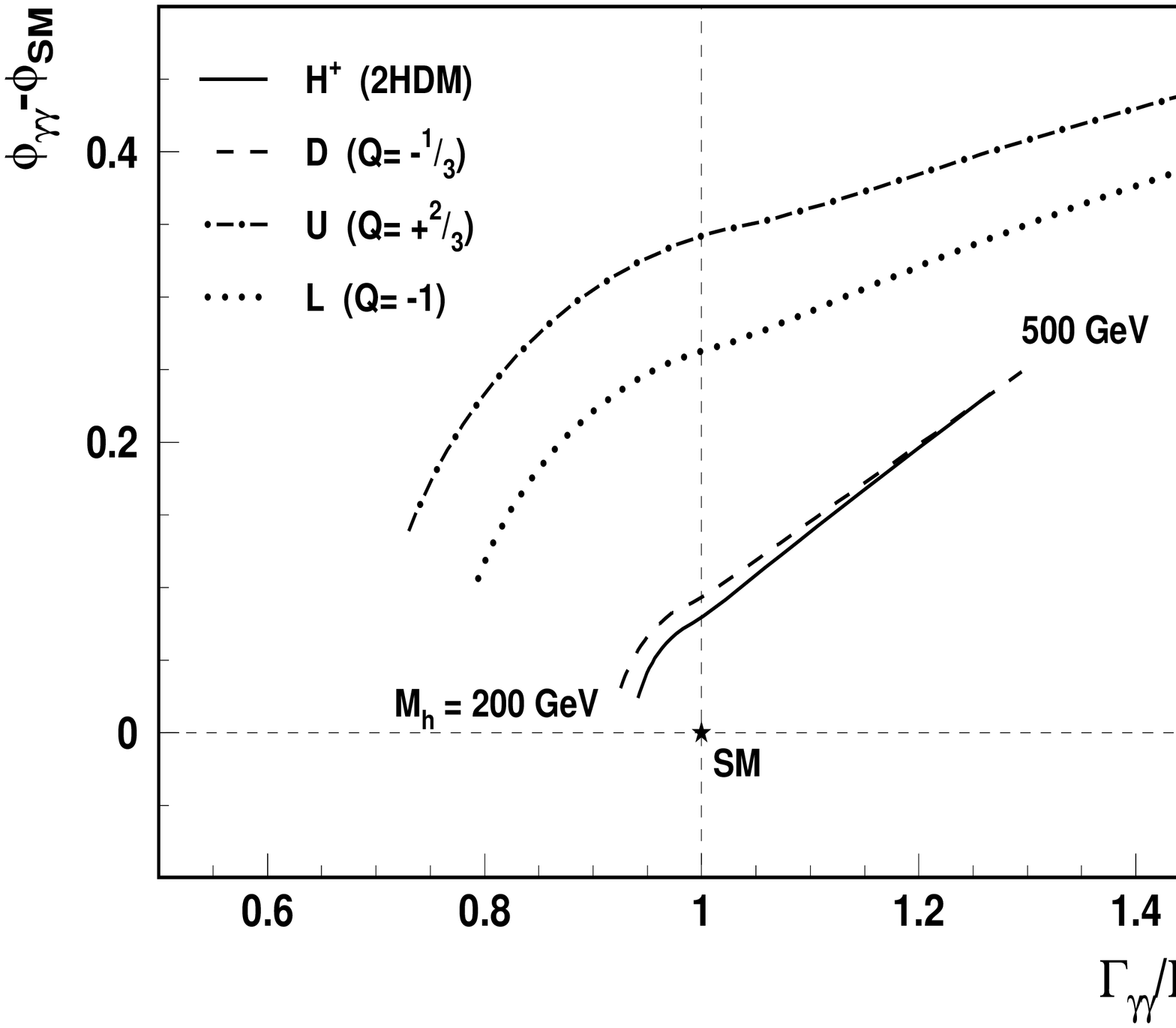}
}
\caption{Results for $\gamma \gamma \ra H,A \ra b \bar b$ in MSSM  for 
``LHC wedge'' (Left and Middle); Right: 
 $\phi_{\gamma \gamma}$ and $\Gamma_{\gamma \gamma}$ in  SM-like models} 
\label{fig:4}       
\end{figure*}

 A resonant production of Higgs boson(s), a unique feature
of PLC, was studied in detail for  Standard Model (SM), MSSM and Two Higgs Doublet Model (2HDM).
\paragraph{$b \bar b$ final state}
The  realistic simulations of the production of SM Higgs boson 
with mass between 120  to 160 GeV decaying into $b\bar b$
were performed \cite{Niezurawski:2003iu,Rosca:2003wa}, including effect of 
overlaying events (OE)
\cite{Niezurawski:2003iu}. The accuracy of extraction of the $\Gamma_{\gamma \gamma} Br (H\ra b \bar b) $ is between 2 to 7 \% (with OE) (Fig.~\ref{fig:3}).
The realistic analysis \cite{Niezurawski:2003ir}
of  production of heavy Higgs bosons $H$ and $A$ 
in MSSM, with parameters \cite{Muhlleitner:2001kw} 
corresponding  to a case where only one SM-like 
Higgs particle $h$ can be seen  at LHC (``LHC wedge''),
 shows large  potential of PLC in search of $H/A$
(Fig.~\ref{fig:4} (Left and Middle)).  

\begin{figure*}
\centering
\resizebox{0.85\textwidth}{!}{%
\includegraphics{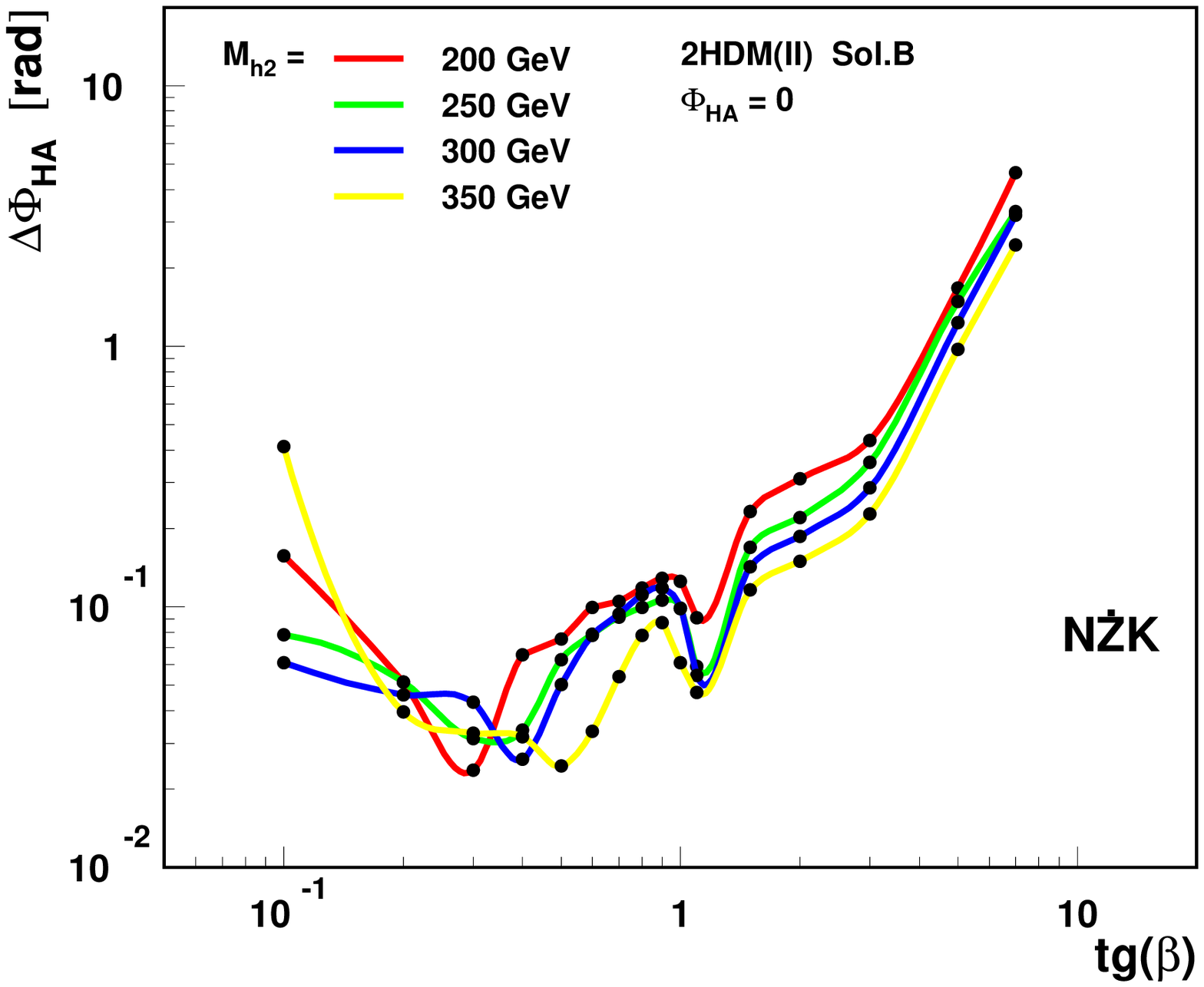}\includegraphics{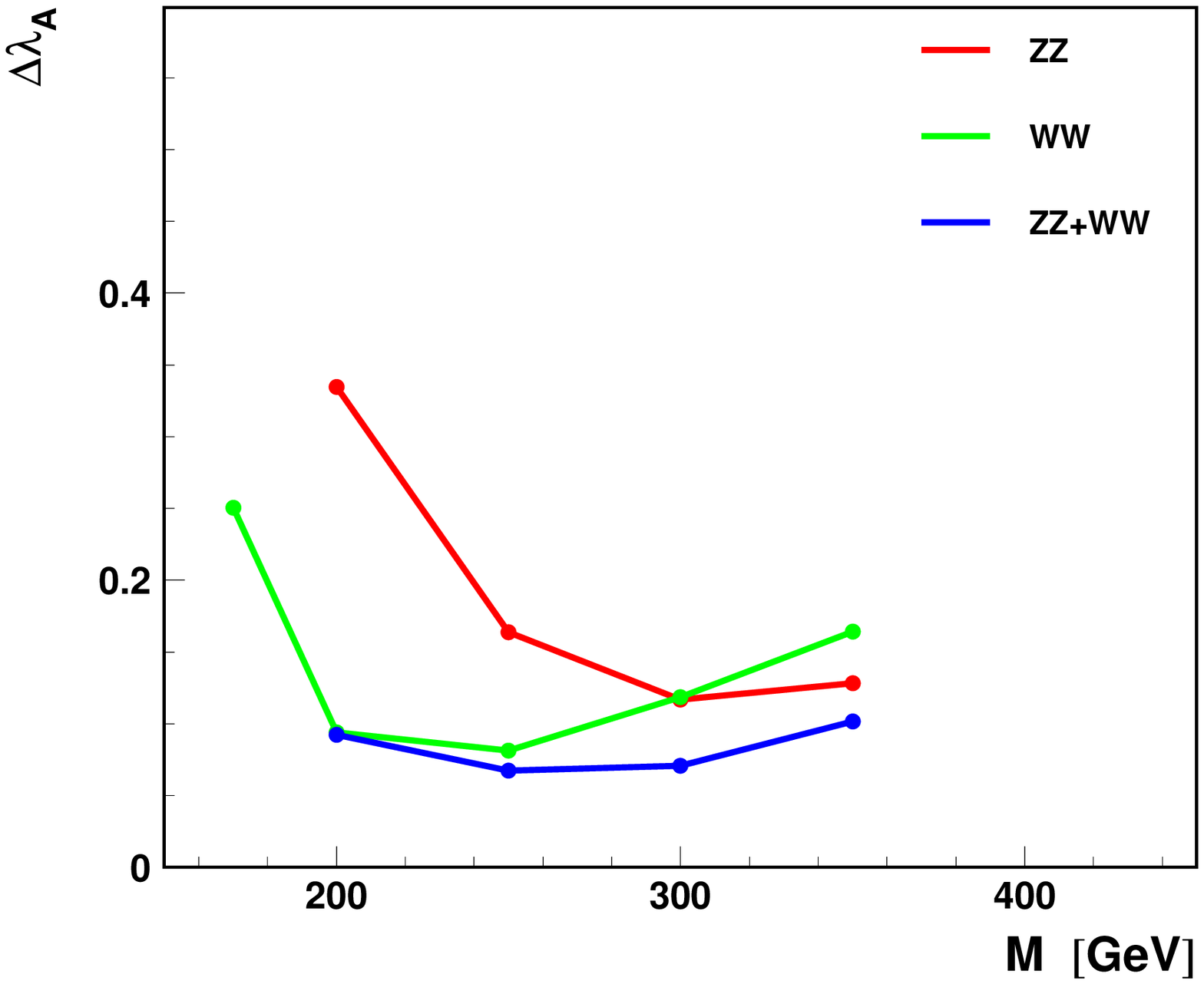}
 \includegraphics{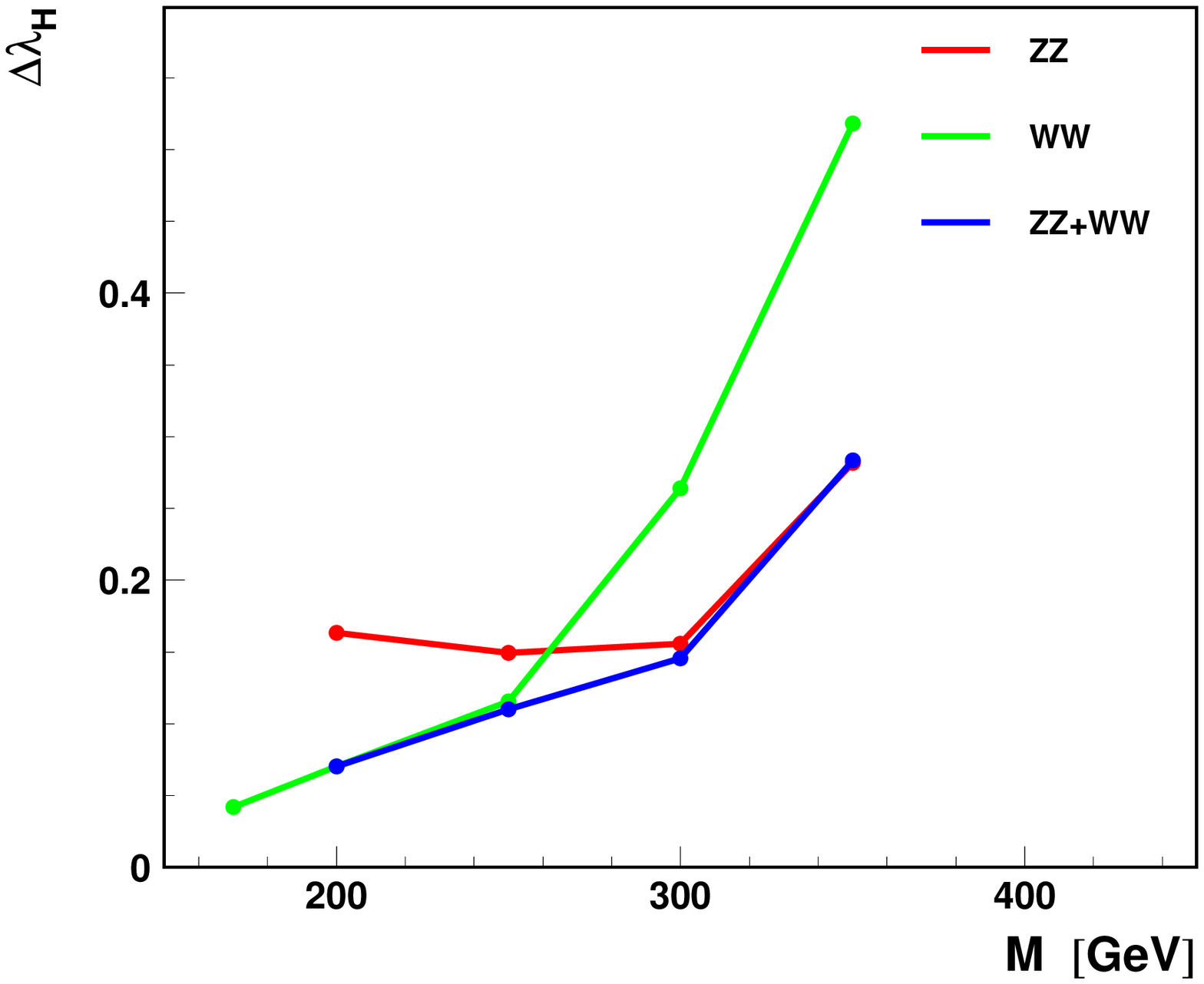}
}
\caption{
Left: The precision of the determination of mixing angle between $H$ and $A$
in 2HDM with small CP-violation; Midle and Right: 
the precision of extraction  of CP-even ($\lambda_H$) and CP-odd ($\lambda_A$) 
coupling to gauge bosons for a generic CP-violation.}
\label{fig:5}
\end{figure*}
\paragraph{WW and ZZ final states}A detailed study of Higgs boson $\phi$,
with or without defined CP-parity, 
 in processes $\gamma \gamma \ra \phi \ra WW/ZZ$ is presented in \cite{Niezurawski:2002jx}.
It was found that interference with background allows to measure besides the 
decay width $\Gamma_{\gamma \gamma}$ also the phase of amplitude $\phi_{\gamma \gamma}$. This enlarges
a discrimination power for  various SM-like extensions
(Fig. ~\ref{fig:4} (Right)),  it is also useful to combine
  WW and ZZ channels. 
Parameters of CP-violation effects can be measured precisely:
mixing angle $\phi_{HA}$ in 2HDM  and couplings $\lambda_{A,H}$ for a generic 
case, shown in Fig.~\ref{fig:5}.
\section{Anomalous gauge coupling in $e\gamma$ collision}
\label{sec:4}
\begin{figure*}
\begin{center}
\resizebox{0.55\textwidth}{!}{%
\includegraphics{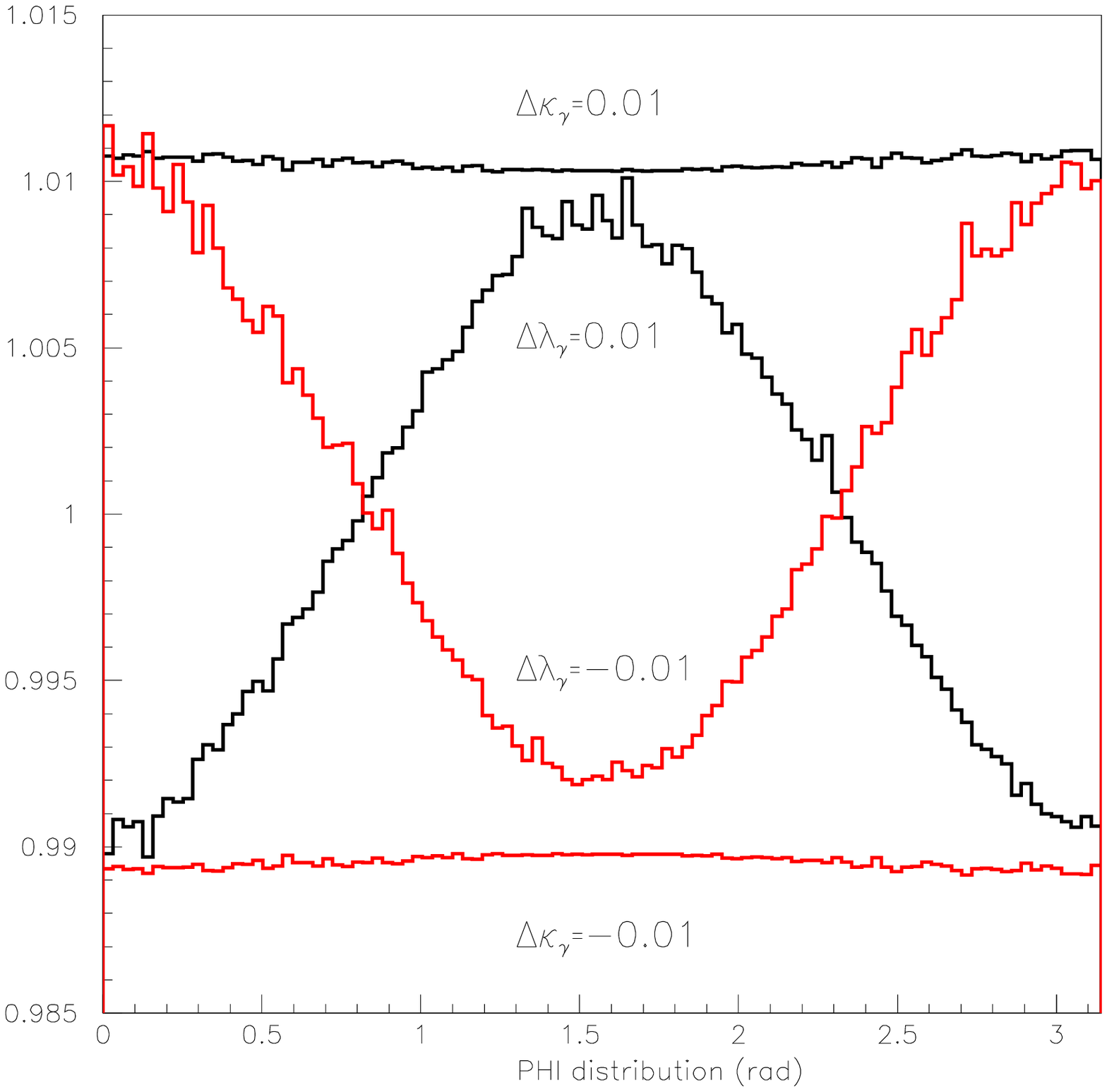}\hspace{1cm}
\includegraphics{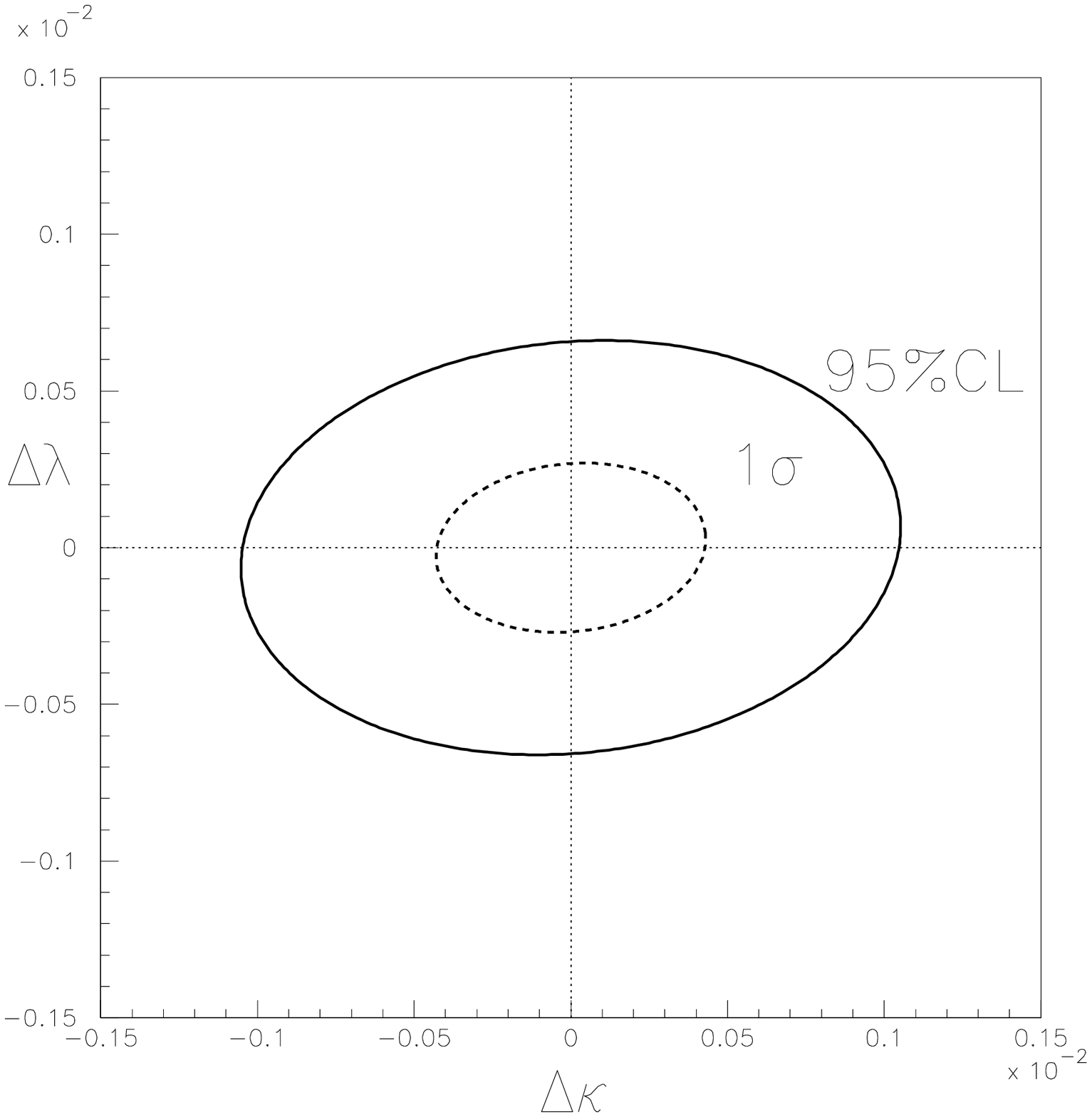}
}
\vspace{-.5in}
\caption{Results for ${\Delta}{\lambda}_{\gamma}$ and ${\Delta}{\kappa}_{\gamma}$. Left: Deviation in $\phi$ distribution; Right: 95$\%$ CL (---) and 1$\sigma$ (- - -) contour plots.}
\label{fig:6}
\end{center}
\end{figure*}

A study of measuring trilinear gauge couplings, 
${\kappa}_{\gamma}$ ${\lambda}_{\gamma}$, from  the hadronic decay of W  at 
an $e\gamma$ - collider at energy 450 GeV was performed in \cite{monig}.
An expected error are  $\sim  10^{-3}$ for ${\kappa}_{\gamma}$ and 
$10^{-4}$ for ${\lambda}_{\gamma}$ if  fit includes the 
azimuthal angle $\phi$ of final fermion (Fig. ~\ref{fig:6}(Left)). 
The contour plot for the deviation from SM  for both couplings is given 
in Fig.\ref{fig:6} (Right). It was found, that 
the uncertainty due to the variable photon beam polarizations is large for ${\kappa}_{\gamma}$, while negligible for ${\lambda}_{\gamma}$. 
\vspace*{-0.3cm}
\section{Outlook}
\label{sec:5}
A new ECFA Study continues  precision theoretical 
and experimental studies of potential of LC for  Higgs search and effects of 
new physics 
for $e^+e^-$ and $\gamma \gamma$ and $e \gamma$ options. \\

{\bf{Acknowledgment:}} I am grateful to K. Desch,P. Nie-\\\.zurawski, 
F. A. \.Zarnecki, K. Moenig and J. Sekaric for valuable 
contributions to this summary.
%
%

\end{document}